%% file: EQUI.tex
\documentclass[12pt]{article}
\usepackage{fullpage,graphicx,color}
\usepackage{amsmath,amssymb,enumerate,epsfig,multirow}
\usepackage[numbers]{natbib}

\newtheorem{lemma}{Lemma}
\newtheorem{theorem}{Theorem}
\newtheorem{proposition}{Proposition}

\newenvironment{myproof}{{\it Proof:\/}}{\hspace*{1em}\hfill$\Box$\vskip 0.1in}
\title{Non-clairvoyant Scheduling Games}
\author{Johanne Cohen%
\thanks{CNRS and PRiSM,
Universit\'e de Versailles St-Quentin-en-Yvelines, France} 
\and Christoph D\"urr%
\thanks{CNRS and LIP6, Universit\'e Pierre et Marie Curie, Paris, France}
\and Nguyen Kim Thang%
\thanks{LAMSADE,
Universit\'e Paris Dauphine, France}
}

\date{}

\begin{document}

\maketitle
\begin{abstract}
In a scheduling game, each player owns a job and chooses a machine to execute it.  While the social cost is the maximal load over all machines (makespan), the cost (disutility) of each player is the completion time of its own job.  In the game, players may follow selfish strategies to optimize their cost and therefore their behaviors do not necessarily lead the game to an equilibrium. Even in the case there is an equilibrium, its makespan might be much larger than the social optimum, and this inefficiency is measured by the price of anarchy -- the worst ratio between the makespan of an equilibrium and the optimum.  Coordination mechanisms aim to reduce the price of anarchy by designing scheduling policies that specify how jobs assigned to a same machine are to be scheduled. 
Typically these policies define the schedule according to the processing times as announced by the jobs. One could wonder if there are policies that do not require this knowledge, and still provide a good price of anarchy. This would make the processing times be private information and avoid the problem of truthfulness.
In this paper we study these so-called \emph{non-clairvoyant} policies.  In particular, we study the \textsf{RANDOM} policy that schedules the jobs in a random order without preemption, and the \textsf{EQUI} policy that schedules the jobs in parallel using time-multiplexing, assigning each job an equal fraction of CPU time.  

For these models we study two important questions, the existence of Nash equilibria and the price of anarchy.  We show that the game under \textsf{RANDOM} policy is a potential game for uniform machines or for two unrelated machines. However, it is not a potential game for three or more
unrelated machines. Moreover, we prove that the game under the \textsf{EQUI} policy is a potential game. 

Next, we analyze the inefficiency of \textsf{EQUI} policy. Interestingly, the (strong) price of anarchy of \textsf{EQUI}, a non-clairvoyant policy, is asymptotically the same as that of the best \emph{strongly local} policy -- policies in which a machine may look at the processing time of jobs assigned to it. The result also indicates that knowledge of jobs' characteristics is not necessarily needed.

\end{abstract}


\section{Introduction}
% lack of coordination & need of design
With the development of the Internet, large-scale autonomous systems became more and more important. The systems consist of many independent and selfish agents who compete for the usage of shared resources.  Every configuration has some social cost, as well as individual costs for every agent.  Due to the lack of coordination, the equilibrium configurations may have high cost compared to the global social optimum and this inefficiency can be captured by the \emph{price of anarchy}~\cite{Koutsoupias.Papadimitriou}.  It is defined as the ratio between the worst case performance of Nash equilibrium~\cite{Nash:Non-Cooperative-Games} and the global optimum.  Since the behavior of the agents is influenced by the individual costs, it is natural to come up with mechanisms that both force the existence of Nash equilibria and reduce the price of anarchy.  The idea is to try to reflect the social cost in the individual costs, so that selfish agents' behaviors result in a socially desired solution.  In particular we are interested in scheduling games, where every player has to choose one machine on which to execute its job.  The individual cost of a player is the completion time of its job, and the social cost is the largest completion time over all jobs, the \emph{makespan}.  For these games, so called \emph{coordination mechanisms} have been studied by~\citet{ChristodoulouKoutsoupiasNanavati:Coordination-Mechanisms}.  A \emph{coordination mechanism} is a set of \emph{local policies}, one for every machine, that specify a schedule for the jobs assigned to it, and the schedule can depend only on these jobs.  Most prior studied policies depend on the processing times and need the jobs to announce their processing times. The jobs could try to influence the schedule to their advantage by announcing not their correct processing times. There are two ways to deal with this issue. One is to design \emph{truthful coordination mechanisms} where jobs have an incentive to announce their real processing times. Another way is to design mechanisms that do not depend on the processing times at all and this is the subject of this paper: we study coordination mechanisms based on so called \emph{non-clairvoyant policies} that we define in this section.

\subsection{Preliminaries}

% Machine scheduling problem, objective function  
\paragraph{Scheduling}
The \emph{machine scheduling problem} is defined as follows: we are given $n$ jobs, $m$ machines and each job needs to be scheduled on exactly one machine.  In the most general case machine speeds are unrelated, and for every job $1\leq i\leq n$ and  every machine $1 \leq j \leq m$ we are given an arbitrary processing time $p_{i,,j}$, which is the time spend by job $i$ on machine $j$.  A schedule $\sigma$ is a function mapping each job to some machine.  The \emph{load} of a machine $j$ in schedule $\sigma$ is the total processing time of jobs assigned to this machine, i.e., $\ell_j = \sum_{i: \sigma(i) = j} p_{i,,j}$.  The \emph{makespan} of a schedule is the maximal load over all machines, and is the social cost of a schedule.  It is NP-hard to compute the \emph{global optimum} even for identical machines, that is when $p_{i,,j}$ does not depend on $j$, see \cite[problem SS8]{GareyJohnson:Computers-and-Intractability;-A-Guide}. We denote by OPT the makespan of the optimal schedule.

% Machine environments
\paragraph{Machine environments}
We consider four different machine environments, which all have their own justification.  The most general environment concerns unrelated machines as defined above and is denoted $R||C_{\max}$.
In the \emph{identical} machine scheduling model, denoted $P||C_{\max}$, every job $i$ comes with a length $p_i$  such that $p_{i,,j}=p_i$ for every machine $j$.  In the \emph{uniform} machine scheduling model, denoted  $Q||C_{\max}$, again every job has \emph{length} $p_i$ and every machine $j$ a speed $s_j$ such that $p_{i,,j}= p_i / s_j$. For the \emph{restricted identical} machine model, every job $i$ comes with a length $p_i$ and a set of machines $S_i$ on which it can be scheduled, such that $p_{i,,j}=p_i$ for $j\in S_i$ and $p_{i,,j}=\infty$ otherwise. In \cite{Brucker:Scheduling-Algorithms} this model is denoted $PMPM||C_{\max}$,  and in \cite{ImmorlicaLiMirrokni:Coordination-Mechanisms-for-Selfish} it is denoted $B||C_{\max}$.

% Selfish scheduling and Nash equilibrium + best-response dynamic + potential game 
\paragraph{Scheduling game}
What we described so far are well known and extensively studied classical scheduling problems.  But now consider the situation where each of the $n$ jobs is owned by an independent agent. In this paper we will sometimes abuse notation and identify the agent with his job.  The agents do not care about the social optimum, their goal is to complete their job as soon as possible.  We consider the situation where each agent can freely decide on which machine its job is to be scheduled. The actual schedule however is not decided by the agents. We rather fix a \emph{policy}, known to all agents, which specifies the actual schedule, once all agents assigned their jobs to machines. Different policies are defined below.

In the paper, we concentrate on \emph{pure strategies} where each agent selects a single machine to process its job.  Such a mapping $\sigma$ is called a \emph{strategy profile}.  Each agent is aware of the decisions made by other agents and behaves selfishly.   The \emph{individual cost} of a job is defined as its completion time.  A \emph{pure Nash equilibrium} is a schedule in which no agent has an incentive to unilaterally switch to another machine. In this paper we will simply omit the adjective \emph{pure}, since there is no confusion possible.  A \emph{strong} Nash equilibrium is a schedule that is resilient to deviations of any coalition, i.e., no group of agents can cooperate and change their strategies in such a way that all players in the group strictly decrease their costs, see~\cite{Aumann:Acceptable-points-in-general,FiatKaplanLevy:Strong-Price-of-Anarchy}.  For some given strategy profile, a \emph{better response move} of a job $i$ is a strategy (machine) $j$ such that if job $i$ changes to job $j$, while all other players stick to their strategy, the cost of $i$ decreases strictly.  If there is such a move, we say that this job is \emph{unhappy}, otherwise it is \emph{happy}.  In this setting a Nash equilibrium is a strategy profile where all jobs are happy.  The \emph{better-response dynamic} is the process of repeatedly choosing an arbitrary unhappy job and changing it to an arbitrary better response move.  A \emph{potential game} is a game in which for any instance, the better-response dynamic always converges~\cite{MondererShapley:Potential-Games}. Such a property is typically shown by the use of a potential function, which maps strategy profiles to non-negative numerical values. The game is called a \emph{strong potential game} if there is a potential function with the property that if an agent improves its individual cost by some amount $\Delta$, then the potential function decreases by the same amount $\Delta$.

% Coordination mechanism, (S)PoA 
A \emph{coordination mechanism} is a set of \emph{scheduling policies}, one for each machine, that determines how to schedule jobs assigned to a machine~\cite{ChristodoulouKoutsoupiasNanavati:Coordination-Mechanisms}. The idea is to connect the individual cost to the social cost, in such a way that  the selfishness of the agents will lead to equilibria that have low social cost.  How good is a given coordination mechanism?  This is measured by the well-known \emph{price of anarchy} (PoA), see~\cite{Koutsoupias.Papadimitriou}.  It is defined as the ratio between the cost of the worst Nash equilibrium and the optimal cost, which is not an equilibrium in general.  We also consider the \emph{strong price of anarchy} (SPoA) which is the extension of the price of anarchy applied to strong Nash equilibria~\cite{FiatKaplanLevy:Strong-Price-of-Anarchy}.

% Policy
\paragraph{Policies} A policy is a rule that specifies how the jobs that are assigned to a machine are to be scheduled. We now define several policies, and give proper credit to the introducing papers in section~\ref{sec:relatedWork}.

%Let $\sigma$ be a strategy profile.
 We distinguish between \emph{local, strongly local} and \emph{non-clairvoyant} policies. Let $S_j$ be the set of jobs assigned to machine $j$. A policy is \emph{local} if the scheduling of jobs on machine $j$ depends only on the parameters of jobs in $S_j$, i.e., it may looks at the processing time $p_{i,k}$ of a job $i \in S_j$ on any machine $k$. A policy is \emph{strongly local} if it looks only at the processing time of jobs in $S_j$ on machine $j$. We call a policy \emph{non-clairvoyant} if the scheduling of jobs on machine $j$ does not depend on the processing time of any job on any machine. 
In this paper we only study coordination mechanisms that use the same policy for all machines, as opposed to \citet{AngelBampisPascual:Truthful-algorithms-for-scheduling}.  \textsf{SPT} and \textsf{LPT} are  policies that schedule the jobs without preemption respectively in order of increasing or decreasing processing times with a deterministic tie-breaking rule for each machine.  An interesting property of \textsf{SPT} is that it minimizes the sum of the completion times, while \textsf{LPT} has a better price of anarchy, because it incites small jobs to go on the least loaded machine which smoothes the loads.  A policy that relates individual costs even stronger to the social cost is \textsf{MAKESPAN}, where jobs are scheduled in parallel on one machine using time-multiplexing and assigned each job a fraction of the CPU that is proportional to its processing time.  As a result all jobs complete at the same time, and the individual cost is the load of the machine. All these policies are \emph{deterministic}, in the sense that they map strategy profiles to a determined schedule. This is opposed to \emph{randomized policies} which map strategy profiles to a distribution of schedules.

What could a scheduler do in the non-clairvoyant case?  He could either schedule the jobs in a random order or in parallel.  The \textsf{RANDOM} policy schedules the jobs in a random order without preemption. Consider a job $i$ assigned to machine $j$ in the schedule $\sigma$, then the cost of $i$ under the \textsf{RANDOM} policy is its expected completion time, i.e., 
$$
c_i = p_{i,j} + \frac12 \sum_{i': \sigma(i') = j,\; i'\neq i} p_{i',j}.
$$
In other words the expected completion time of $i$ is half of the total load of the machine, where job $i$ counts twice.  Again, as for \textsf{MAKESPAN}, the individual and social cost in \textsf{RANDOM} are strongly related, and it is likely that these policies should have the same price of anarchy.  That is is indeed the case except for unrelated machines.  

  \begin{figure}[tbp]
    \centering{
     \includegraphics[width=12cm]{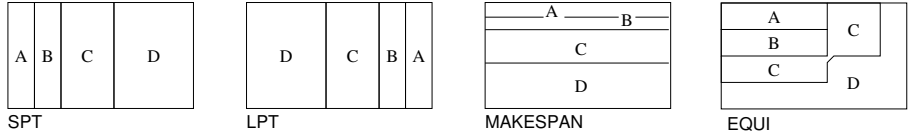}
    }
    \caption{Different scheduling policies for $p_A=1, p_B=1, p_C=2, p_D=3$. Tie is broken arbitrarily between jobs $A$ and $B$. The rectangles represent the schedules on a single machine with time going from left to right and the height of a block being the amount of CPU assigned to the job.}
    \label{fig:policy}
  \end{figure}

Another natural non-clairvoyant policy is \textsf{EQUI}, which has been studied for example in~\cite{Edmonds:Scheduling-in-the-dark} in the different context of online algorithms.  As \textsf{MAKESPAN} it schedules the jobs in parallel preemptivly using time-multiplexing, but it assigns to every job the same fraction of the CPU.   Suppose there are $k$ jobs with processing times $p_{1,j} \leq p_{2,j} \leq \ldots \leq p_{k,j}$ assigned to machine $j$, we renumbered jobs from $1$ to $k$ for this example.  Since, each job receives the same amount of resource, then job $1$ is completed at time $c_1=kp_{1,j}$. At that time, all jobs have remaining processing time $(p_{2,j} - p_{1,j}) \leq (p_{3,j} - p_{1,j}) \leq \ldots \leq  (p_{k,j} - p_{1,j})$. Now the machine splits its resource into $k-1$ parts until the moment job 2 is completed, which is at $kp_{1,j} + (k-1)(p_{2,j} - p_{1,j}) = p_{1,j} + (k-1)p_{2,j}$. In general, the completion time of job $i$, which is also its cost, under \textsf{EQUI} policy is:
\begin{align}
c_i &= c_{i-1} + (k-i+1)(p_{i,j} - p_{i-1,j})  \notag \\
 &= p_{1,j} + \ldots + p_{i-1,j} + (k-i+1)p_{i,j}	\label{eq:EQUI}
\end{align}

We already distinguished policies depending on what information is needed from the jobs.  In addition we distinguish between \emph{preemptive} and \emph{non-preemptive} policies, depending on the schedule that is produced.  Among the policies we considered so far, only \textsf{MAKESPAN} and \textsf{EQUI} are preemptive, in the sense that they rely on time-multiplexing, which consists in executing arbitrary small slices of the jobs. Note that, \textsf{EQUI} is a realistic and quite popular policy. It is implemented in many operating systems such as Unix and Windows. See Figure~\ref{fig:policy} for an illustration of these five policies.

\paragraph{Example}
For illustration consider the scheduling game on parallel identical machines and the \textsl{EQUI} policy. Here each of the $n$ jobs has a processing time $p_i$, for $1\leq i\neq n$. Every agent selects a machine, which is described by a strategy profile $\sigma : \{1,\ldots,n\} \rightarrow \{1,\ldots,m\}$. Now the individual cost of agent $i$, is the completion time of its job, which for this policy is 
\[
       \sum_{i'} \min\{p_{i,j},p_{i',j}\},
\]
 where the sum is taken over all jobs $i'$ assigned to the same machine as $i$, i.e.~$\sigma(i)=\sigma(i')$.

\subsection{Previous and related work}    \label{sec:relatedWork}

Coordination mechanism are related to local search algorithms. The local improvement moves in the local search algorithm correspond to the better-response moves of players in the game defined by the coordination mechanism. Some results on local search algorithms for scheduling problem are surveyed in~\cite{Vredeveld:Combinatorial-Approximation-Algorithms:}.

Most previous work concerned non-preemptive strongly local policies, in particular the \textsf{MAKESPAN} policy. \citet{CzumajVocking:Tight-bounds-for-worst-case} gave tight results $\Theta(\log m/\log \log m)$ of its price of anarchy for pure Nash equilibria on uniform machines. \citet{FiatKaplanLevy:On-the-Price-of-Stability-for-Designing} extended this result for the strong price of anarchy, and obtained the tight bound $\Theta(\log m / (\log \log m)^2)$. In addition,~\citet{GairingLuckingMavronicolas:Computing-Nash-equilibria} and~\citet{AwerbuchAzarRichter:Tradeoffs-in-worst-case-equilibria} gave tight bounds for the price of anarchy for restricted identical machines.

Coordination mechanism design was introduced by~\citet{ChristodoulouKoutsoupiasNanavati:Coordination-Mechanisms}.  They studied the \textsf{LPT} policy on identical machines.  \citet{ImmorlicaLiMirrokni:Coordination-Mechanisms-for-Selfish} studied coordination mechanism for all four machine environments and gave a survey on the results for non-preemptive strongly local policies.  They also analyzed the existence of pure Nash equilibria under \textsf{SPT}, \textsf{LPT} and \textsf{RANDOM} for certain machine environments and the speed of convergence to equilibrium of the better response dynamics.  Precisely, they proved that the game is a potential game under the policies \textsf{SPT} on unrelated machines, \textsf{LPT} on uniform or restricted identical machines, and \textsf{RANDOM} on restricted identical machines.  
In~\cite{Thang:Phd} it was shown that the game does not converge under the \textsf{LPT} policy on unrelated machines.
The policy \textsf{EQUI} has been studied in \cite{Edmonds:Scheduling-in-the-dark} for its competitive ratio.
The results are summarized in Table~\ref{table:strongly-local policy}.

\citet{AzarJainMirrokni:Almost-Optimal-Coordination} introduced the inefficiency-based local policy which has price of anarchy $O(\log m)$ on unrelated machines. Moreover, they also proved that every non-preemptive strongly local policy with an additional assumption has price of anarchy at least $m/2$, which shows a sharp difference between strongly local and local policies.  

\begin{table}[h]		
  \centering
  \begin{tabular}{|l|l|l|l|l|l|}
    \hline
    model $\backslash$ policy & \textsf{MAKESPAN}  & \textsf{SPT} & \textsf{LPT} 
                                                                                              & \textsf{RANDOM}
&\textsf{EQUI} \\
    \hline
    identical & $2 - \frac{2}{m + 1}$
    & $2 - \frac{1}{m}$
    & $\frac{4}{3} - \frac{1}{3 m}$
    & $2 - \frac{2}{m + 1}$  
    & $2 - \frac1m$ 
\\
    & \multicolumn{1}{r|}{\cite{FinnHorowitz:A-linear-time-approximation, SchuurmanVredeveld:Performance-guarantees-of-local}}
    & \multicolumn{1}{r|}{\cite{Graham:Bounds-for-certain-multiprocessing, ImmorlicaLiMirrokni:Coordination-Mechanisms-for-Selfish}}
    & \multicolumn{1}{r|}{\cite{Graham:Bounds-on-multiprocessing-timing, ChristodoulouKoutsoupiasNanavati:Coordination-Mechanisms}}
    & \multicolumn{1}{r|}{\cite{FinnHorowitz:A-linear-time-approximation, SchuurmanVredeveld:Performance-guarantees-of-local}}
&
                 \\
    \hline
    uniform & $\Theta ( \frac{\log m}{\textrm{loglog} m})$  
    & $\Theta (\log m)$  
    & $1.52 \leq PoA \leq 1.59$ 
    & $\Theta ( \frac{\log m}{\textrm{loglog} m})$ 
    & $\Theta ( \log m )$
\\
    & \multicolumn{1}{r|}{\cite{CzumajVocking:Tight-bounds-for-worst-case}}
    	& \multicolumn{1}{r|}{\cite{AspnesAzarFiat:On-line-routing-of-virtual, ImmorlicaLiMirrokni:Coordination-Mechanisms-for-Selfish}}
    	& \multicolumn{1}{r|}{\cite{Dobson:Scheduling-independent-tasks, Friesen:Tighter-bounds-for-LPT-scheduling, ImmorlicaLiMirrokni:Coordination-Mechanisms-for-Selfish}}
	& \multicolumn{1}{r|}{\cite{CzumajVocking:Tight-bounds-for-worst-case}} 
        &
\\
    \hline
    restricted id. & $\Theta ( \frac{\log m}{\textrm{loglog} m})$ 
     
    & $\Theta (\log m)$ 
     
    & $\Theta (\log m)$ 
   
    & $\Theta ( \frac{\log m}{\textrm{loglog} m})$ 
    &  $\Theta (\log m)$
     \\
        & \multicolumn{1}{r|}{\cite{GairingLuckingMavronicolas:Computing-Nash-equilibria,AwerbuchAzarRichter:Tradeoffs-in-worst-case-equilibria}}
        & \multicolumn{1}{r|}{\cite{AspnesAzarFiat:On-line-routing-of-virtual, ImmorlicaLiMirrokni:Coordination-Mechanisms-for-Selfish}}
        & \multicolumn{1}{r|}{\cite{AzarNaorRom:The-Competitiveness-of-On-Line-Assignments, ImmorlicaLiMirrokni:Coordination-Mechanisms-for-Selfish}}
        & \multicolumn{1}{r|}{\cite{GairingLuckingMavronicolas:Computing-Nash-equilibria,AwerbuchAzarRichter:Tradeoffs-in-worst-case-equilibria}}
&
\\
    \hline
    unrelated & unbounded
    & $\Theta (m)$
    & unbounded 
    & $\Theta ( m)$
    & $\Theta ( m)$
\\
    & \multicolumn{1}{r|}{\cite{SchuurmanVredeveld:Performance-guarantees-of-local}}
    & \multicolumn{1}{r|}{\cite{ChoSahni:Bounds-for-list-schedules, IbarraKim:Heuristic-algorithms-for-scheduling, AzarJainMirrokni:Almost-Optimal-Coordination}}
    & \multicolumn{1}{r|}{}
    & \multicolumn{1}{r|}{\cite{ImmorlicaLiMirrokni:Coordination-Mechanisms-for-Selfish}}
    &
\\    \hline
  \end{tabular}
  \caption{Price of anarchy under different strongly local and non-clairvoyant policies. 
    The right most column is our contribution.}
  \label{table:strongly-local policy}
\end{table}

\subsection{Our contribution}

We are interested in \emph{admissible} non-clairvoyant policies -- policies that always induce a Nash equilibrium for any instance of the game.
In the game, maybe more important than the question of existence of Nash equilibrium is the question of convergence to an equilibria. Since no processing time is known to the coordination mechanism it is impossible to compute some equilibria or even decide if a given assignment of jobs to machines is an equilibria. Besides, if all processing times are known to all jobs, it makes sense to let the jobs evolve according to the better-response dynamics, until they eventually reach an equilibria. Therefore in the paper, we are interested in the convergence of the better-response dynamic. 

 In Section~\ref{section:existence}, we study the existence of Nash equilibrium under the non-clairvoyant policies \textsf{RANDOM} and \textsf{EQUI}.  We show that for the \textsf{RANDOM} policy, the game is a potential game on uniform machines.  We also show that on two unrelated machines, it is a potential game, but for three unrelated machines or more, the better-response dynamic does not converge. Moreover, we prove that for the \textsf{EQUI} policy, the game is a (strong) potential game, see Table~\ref{table:convergence}.

\begin{table}[h]		
  \centering
  \begin{tabular}{|l|l|l|l|l|l|}
    \hline
    model $\backslash$ policy & \textsf{MAKESPAN}  & \textsf{SPT} & \textsf{LPT} 
                                                                                              & \textsf{RANDOM}
&\textsf{EQUI} \\
    \hline
    identical & \multirow{5}{4em}{Yes\\{\cite{Even-DarKesselmanMansour:Convergence-time-to-Nash}}} &
    \multirow{5}{3em}{Yes\\ \cite{ImmorlicaLiMirrokni:Coordination-Mechanisms-for-Selfish}} &
    \multirow{4}{3em}{Yes\\ \cite{ImmorlicaLiMirrokni:Coordination-Mechanisms-for-Selfish}}&
    Yes \cite{ImmorlicaLiMirrokni:Coordination-Mechanisms-for-Selfish}&
    \multirow{5}{3em}{Yes} 
    \\ \cline{1-1}\cline{5-5}
    uniform &&& & Yes &\\\cline{1-1}\cline{5-5}
    restricted id. &&& & Yes \cite{ImmorlicaLiMirrokni:Coordination-Mechanisms-for-Selfish} &\\\cline{1-1}\cline{4-5}
    \multirow{2}{7em}{unrelated} &&& \multirow{2}{3em}{No~\cite{Thang:Phd}} & Yes for $m=2$ & \\ 
\cline{5-5}
    &&& & No for $m\geq 3$ & \\ \hline
    \end{tabular}
\caption{Convergence of the better response dynamic.}
\label{table:convergence}
\end{table}

In Section~\ref{section:inefficiency}, we analyze the price of anarchy and the strong price of anarchy of \textsf{EQUI}. We observe that \textsf{RANDOM} is slightly better than \textsf{EQUI} except for the unrelated model.  In the unrelated model, interestingly, the price of anarchy of \textsf{EQUI} reaches the lower bound in \cite{AzarJainMirrokni:Almost-Optimal-Coordination} on the PoA of any strongly local policy with some additional condition.  The latter shows that although there is a clear difference between strongly local and local policies with respect to the price of anarchy, our results indicate that in contrast, restricting strongly local policies to be non-clairvoyant does not really affect the price of anarchy. Moreover, \textsc{EQUI} policy does not need any knowledge about jobs' characteristics, even their identities (IDs) which are useful in designing policies with low price of anarchy in \cite{AzarJainMirrokni:Almost-Optimal-Coordination, Caragiannis:Efficient-coordination-mechanisms}.

\section{Existence of Nash equilibrium}		\label{section:existence}
The results in this section are summarized as follows.
\\  
 
\noindent 
\textbf{Summary of results on the existence of Nash equilibrium:}
{ \it  We consider the scheduling game under different policies in different machine environments.
  \begin{enumerate}

    \item For the \textsf{RANDOM} policy on uniform machines, it is a potential game.
    For the \textsf{RANDOM} policy on unrelated machines, it is not a potential game for 3 or more machines, 
    but it is a potential game for 2 machines.

    \item For the \textsf{EQUI} policy it is an exact potential game.
  \end{enumerate}
}

%%%%%%%%%%%%%%%%%%%%%%%%%%%%%
%  RANDOM in Q||C_max, existence
%%%%%%%%%%%%%%%%%%%%%%%%%%%%%

\subsection{The \textsf{RANDOM}  policy  on uniform machines}	\label{subsection:RANDOM_uniform}
In the \textsf{RANDOM} policy, the cost of a job is its expected completion time. If the load of machine $j$ is $\ell_j$ then the cost of job $i$ assigned to machine $j$ is $\frac12(\ell_j+p_{i,j})$.  Observe that a job $i$ on machine $j$ has an incentive to move to machine $j'$ if and only if $p_{i,j} + \ell_j > 2p_{i,j'} + \ell_{j'}$. 

In this section, we consider uniform machines. 
Let $p_1 \leq p_2 \leq \ldots \leq p_n$ be the job lengths and 
$s_1 \geq s_2 \geq \ldots \geq s_m$ be the machine speeds. 
Now the processing time of job $i$ on machine $j$ is $p_i/s_j$.  

\begin{theorem}
The scheduling game on uniform machines and the \textsf{RANDOM} policy is a potential game.
\end{theorem}   
\begin{myproof}
Let $\sigma:\{1,\ldots,n\}\rightarrow\{1,\ldots,m\}$ be a strategy profile. 
The proof will use a potential function adapted from previous studies on congestion games \cite{Rosenthal:A-Class-of-Games-Possessing} and load balancing games \cite{GoldbergPODC2004,Fotakis05}. 
We define
\[
   \Phi(\sigma) := \sum_{j=1}^m \frac{\ell_j^2}{s_j} + 3 \sum_{i=1}^n \frac{p_i^2}{s_{\sigma(i)}},
\]
where $\ell_j$ is the load of machine $j$, i.e.\ the sum of $p_i$ over all jobs $i$ with $\sigma(i)=j$.

Now consider a job $i$ that makes a better response move from machine $a$ to machine $b$. If $\ell_a,\ell_b$ denote the loads respectively of machine $a$ and $b$ \emph{before} the move, then by definition of a better response move we have the inequality
\begin{equation} \label{eq:happy}
   \frac{\ell_a + p_i}{s_a} > \frac{ \ell_b + 2 p_i }{s_b}.
\end{equation}

Let $\sigma'$ be the profile \emph{after} the move of job $j$. The change in the potential is
\begin{align*}
\Phi(\sigma') - \Phi(\sigma)
 &= \frac{(\ell_b+p_i)^2}{s_b} + \frac{(\ell_a-p_i)^2}{s_a} + \frac{3 p_i^2}{s_b} - 
       \frac{\ell_a^2}{s_a} - \frac{\ell_b^2}{s_b} - \frac{3 p_i^2}{s_a}
\\
&= \frac{\ell_b^2 +2 \ell_b p_i + p_i^2 + 3 p_i^2 - \ell_b^2}{s_b} 
+    \frac{\ell_a^2 - 2\ell_a p_i + p_i^2 - \ell_a^2 - 3p_i^2}{s_a} 
\\
&=
    2 p_i \left( \frac{\ell_b + 2p_i}{s_b} - \frac{\ell_a+p_i}{s_a} \right) < 0
\end{align*}
due to (\ref{eq:happy}).
Therefore, the potential function $\Phi$ strictly decreases at every better response move.
\end{myproof}

%%%%%%%%%%%%%%%%%%%%%%%%%%%%%
%  RANDOM potential game for 2 machines but not > 2
%%%%%%%%%%%%%%%%%%%%%%%%%%%%%
\subsection{The \textsf{RANDOM} policy for unrelated machines}	\label{subsection:RANDOM_unrelated}
In the following, we will characterize the game under the \textsf{RANDOM} policy in the unrelated model 
as a function of the number of machines. 

\begin{theorem}		\label{theorem:Random2}
The scheduling game on 2 unrelated machines with the \textsf{RANDOM} is a potential game.
\end{theorem}

\begin{myproof}
Let  $\sigma:\{1,\ldots,n\} \rightarrow \{1,2\}$ be the current strategy
  profile, meaning that job $i$ is assigned to machine $\sigma(i)$. By
  $\overline{\sigma}(i)$ we denote the opposite machine to machine $\sigma(i)$.  Let $\ell_j$ be the  load of  machine $j$ in strategy profile $\sigma$, which is  $\sum_{i: \sigma(i) = j} p_{i,j}$.  Define the potential function as
\[
	\Phi(\sigma) := (\ell_{1}-\ell_{2})^{2} + 3 \sum_{i=1}^{n} p_{i,\sigma(i)}^{2}. 
\]

We claim that the potential function $\Phi$ strictly decreases at every better response move.  Let $i$ be a
job moving from say machine $a$ to machine $b$, while strictly decreasing its cost, i.e.
\begin{equation}				\label{eq:move}
 \ell_{b} + 2p_{i,b} - \ell_{a} - p_{i,a} < 0,
\end{equation}
where $\ell_{a},\ell_{b}$ are the loads before the move. 

Let $\sigma'$ be the strategy profile after the move of job $i$. We have:
\begin{align*}
\Phi(\sigma') - \Phi(\sigma) 
	&=  (\ell_{a}- p_{i,a} - \ell_{b} - p_{i,b})^{2} - (\ell_{a}-\ell_{b})^{2} + 3(p_{i,b}^{2} - p_{i,a}^{2}) \\
	&=  -(p_{i,a} + p_{i,b}) (2\ell_{a}- p_{i,a} - 2\ell_{b} - p_{i,b}) + 3 (p_{i,a} + p_{i,b}) (p_{i,a} - p_{i,b}) \\
	&= (p_{i,a} + p_{i,b})[3(p_{i,b} - p_{i,a}) - (2\ell_{a}- p_{i,a} - 2\ell_{b} - p_{i,b})] \\
	&= 2(p_{i,a} + p_{i,b})[(2p_{i,b} + \ell_{b}) - (p_{i,a} +\ell_{a})] < 0 
\end{align*}
due to (\ref{eq:move}).
Therefore, the potential function $\Phi$ strictly decreases at every better response move.
\end{myproof}

However, for 3 or more machines, the better-response dynamic does not necessarily converge. 

\begin{lemma}		\label{lem:Random3}
The better-response dynamic does not converge under the \textsf{RANDOM} policy on 3 or more unrelated machines.
\end{lemma}
\begin{myproof}
  We give a simple four-job instance, with the following processing
  times. For convenience we name the jobs $A,B,C,D$.
  \[
  \begin{array}{c|rrr}
    p_{i,j} &   1 &   2 &   3 \\ \hline
    A    &  90 &  84 &\infty \\
    B    &  96 &   2 &\infty \\
    C    & 138 & 100 &\infty \\
    D   &\infty& 254 &   300 \\
  \end{array}
  \]
  Now we describe a cyclic sequence of better response moves, where each job strictly decreases
  its cost, showing that the game does not converge. In the following table, we describe in each line, the current strategy profile, a better response move of an unhappy job and its cost improvement. For example the first line shows the strategy profile, where jobs $A,B$ are on machine $1$, $C$ is on machine $2$ and $D$ on machine $3$. Then job the cost of job $A$ is $138$ and moving to machine $2$, its cost drops to $134$. The subsequent line show similar better response moves, which end in the initial strategy profile.
  \[
  \begin{array}{llllc}
    1&2&3 & \text{move}&\text{cost improvement}\\ \hline
    AB&C&D & 1\stackrel{A}{\rightarrow}2 & 138>134 \\
    B&AC&D & 1\stackrel{B}{\rightarrow}2 &  96> 94\\
    &ABC&D & 2\stackrel{C}{\rightarrow}1 & 143>138\\
    C&AB&D & 3\stackrel{D}{\rightarrow}2 & 300>297\\
    C&ABD& & 2\stackrel{B}{\rightarrow}1 & 171>165\\
    BC&AD& & 2\stackrel{A}{\rightarrow}1 & 211>207\\
    ABC&D& & 1\stackrel{C}{\rightarrow}2 & 231>227\\
    AB&CD& & 2\stackrel{D}{\rightarrow}3 & 304>300\\
    AB&C&D 
  \end{array}
  \]
\end{myproof}

Note that although there exists a cycle in better-reponse dynamic of the game under \textsf{RANDOM} policy, this does not mean that the game possesses no equilibrium, see~\cite{MondererShapley:Potential-Games}. %In fact, this question remains open. 

%%%%%%%%%%%%%%%%%%%%%%%%%%%%%
%  EQUI's convergence
%%%%%%%%%%%%%%%%%%%%%%%%%%%%%

\subsection{The \textsf{EQUI} policy}	\label{subsection:EQUI}

In the \textsf{EQUI} policy, the cost of job $i$ assigned to machine $j$ is given by expression $(\ref{eq:EQUI})$. Here is an alternative formulation for the cost
 $$
 c_i =  \sum_{\begin{subarray}{c} i': \sigma(i') = j \\ 
 								p_{i',j} \leq  p_{i,j} 
 				\end{subarray} } p_{i',j} + 
 				 \sum_{\begin{subarray}{c} i': \sigma(i') =  j\\ 
 								p_{i',j} >  p_{i,j} 
 				\end{subarray} } p_{i,j} 
 $$

\begin{lemma} \label{lem:EQUI-potential}
The game with the \textsf{EQUI} policy is an exact potential game. 
\end{lemma}

\begin{myproof}
If in a game every better response move would strictly decrease the total load, the game would converge. Unfortunately the game does not have this property. Also, if a better response move would never increase the individual costs of players, again the game would converge, since the total individual costs would measure convergence. It happens that the game does not have this property either.
It turns out that a measure for the convergence is in fact an average of two measures above: the sum over all individual costs and the total load over all machines.  
 
Let $\sigma$ be the current strategy profile, meaning $\sigma(i)$ is the current machine on which job $i$ is scheduled.  Consider the following potential function.
$$
\Phi(\sigma) = \frac{1}{2} \sum_{i=1}^{n} \left ( c_i + p_{i,\sigma(i)} \right ) 
$$

We prove that if a job makes a better response move then the potential function strictly decreases.  Let $t$ be a job that moves from  machine $a$ to $b$, while stricly decreasing its cost from $c_{t}$ to $c'_{t}$.  We have
\begin{align*}
c_t &= \left ( \sum_{\begin{subarray}{c} i: \sigma(i) = a, i \neq t \\ 
								p_{i,a} \leq p_{t,a} 
				\end{subarray} } p_{i,a} + 
				 \sum_{\begin{subarray}{c} i: \sigma(i) = a, i \neq t \\ 
								p_{i,a} > p_{t,a} 
				\end{subarray} } p_{t,a} \right ) + p_{t,a}
\\ &> 
  \left ( \sum_{\begin{subarray}{c} i: \sigma(i) = b, i \neq t \\ 
								p_{i,b} \leq p_{t,b} 
				\end{subarray} } p_{i,b} + 
				 \sum_{\begin{subarray}{c} i: \sigma(i) = b, i \neq t \\ 
								p_{i,b} > p_{t,b} 
				\end{subarray} } p_{t,b} \right ) + p_{t,b} 
\\ &= c'_t.
\end{align*}
Let $\sigma'$ be the strategy profile after the move of job $t$.
Note that in $\sigma'$ the processing time of all jobs except $i$ and the cost of all jobs scheduled on machine different to $a$ and $b$ stay the same. Thus, the change in the potential depends only on the jobs scheduled on machines $a$ and $b$. 
\begin{align*}
2 \cdot \Delta \Phi =&
\left (  \sum_{i: \sigma'(i) = a} (c'_i + p_{i,a})  ~+  \sum_{i: \sigma'(i) = b, i \neq t} (c'_i + p_{i,b})  + p_{t,b} \right )
\\
& ~- 
\left (  \sum_{i: \sigma(i) = a, i \neq t} (c_i + p_{i,a})  ~+  \sum_{i: \sigma(i) = b} (c_i + p_{i,b})  ~+ p_{t,a} \right ) + (c'_t - c_t)
\\
=& \sum_{i: \sigma(i) = a, i \neq t} (c'_i - c_i) ~+ \sum_{i: \sigma(i) = b, i \neq t} (c'_i - c_i) ~+ (c'_t - c_t) + p_{t,b} - p_{t,a}
\end{align*}   
since $\sigma(i) = \sigma'(i) ~\forall i \neq t$.

Consider a job $i \neq t$ on machine $a$.  If the processing time of $i$ is at most that of $t$ then the difference between its new and old cost is exactly $-p_{i,a}$.  Otherwise if the processing time of $i$ is strictly greater than that of $t$ then this difference is exactly $-p_{t,a}$.  Analogously for jobs on machine $b$. Hence,
\begin{align*}
2 \cdot \Delta \Phi =& \left ( \sum_{\begin{subarray}{c} i: \sigma(i) = b, i \neq t \\ 
								p_{i,b} \leq p_{t,b} 
				\end{subarray} }  p_{i,b} ~+ 
				 \sum_{\begin{subarray}{c} i: \sigma(i) = b, i \neq t \\ 
								p_{i,b} > p_{t,b} 
				\end{subarray} }  p_{t,b} ~+ p_{t,b} \right ) + \\
		    &+  \left ( \sum_{\begin{subarray}{c} i: \sigma(i) = a, i \neq t \\ 
								p_{i,a} \leq p_{t,a} 
				\end{subarray} } -p_{i,a} ~+ 
				 \sum_{\begin{subarray}{c} i: \sigma(i) = a, i \neq t \\ 
								p_{i,a} > p_{t,a} 
				\end{subarray} } -p_{t,a} ~- p_{t,a} \right ) 
		    + (c'_t - c_t) \\
		    =& ~ 2\cdot (c'_t - c_t) < 0
\end{align*}
Therefore, the game with the \textsf{EQUI} policy is an exact potential game.
\end{myproof}

Now we strengthen the statement of the previous lemma.

\begin{theorem} \label{thm:EQUI-potential}
The game with the \textsf{EQUI} policy is a strong potential game, in the sense that the better-response dynamic converges even with deviations of coalitions.
\end{theorem}
\begin{myproof}
Let $S$ be a coalition and define its total cost $c(S) := \sum_{i \in S} c_i$.  We study a better response move of $S$ by dividing the process into two phases: in the first phase, all jobs in $S$ move out (disappear) from the game and in the second phase, jobs from $S$ move back (appear) into the game at their new strategies.  We argue that after the first phase, the change in the potential is $\Delta \Phi = -c(S)$ and after the second phase $\Delta \Phi = c'(S)$.  Since the argument is the same, we only prove it for the first phase; the second phase can be done similarly.  Fix a machine $a$ and suppose without loss of generality that all the jobs assigned to $a$ are $1,\ldots,k$ for some $k$.  Also to simplify notation we denote $q_{i}=p_{i,a}$ and assume $q_1 \leq \ldots \leq q_k$. Let $R = S\cap \sigma^{-1}(j)= \{i_1 \leq \ldots \leq i_r\}$ be the set of jobs in the coalition that are scheduled on this machine. Then,
\begin{align*}
c(R) = \sum_{j=1}^r c_{i_j} = \sum_{j=1}^r \left ( q_1 + q_2 + \ldots + q_{i_j - 1} + (k - i_j + 1)q_{i_j}  \right )
\end{align*} 

The jobs in $R$ partition the jobs $\{1,\ldots,k\}$ into $r+1$ parts: part $j\in\{0,\ldots,r\}$ is $[i_{j}+1,j_{j+1}]$, where for convenience we denote $i_0 = 0$ and $i_{r+1} = k$. After the move out of $R$, the change in cost of a job $t \notin R$ scheduled on the machine with index in $[ i_j+1, i_{j+1}]$ is $q_{i_1} + q_{i_2} + \ldots + q_{i_{j-1}} + (r-j)q_t$.  Hence, the difference in the potential restricted to machine $a$ after the first phase $ \Delta \Phi |_{a}$ satisfies:
\begin{align*}
-2 \Delta \Phi |_{a} = & \left [ \sum_{j=0}^{r}
\sum_{\substack{ t \notin R \\ t \in [ i_j + 1, i_{j+1}]} } q_{i_1} + q_{i_2} + \ldots + q_{i_{j-1}} + (r-j)q_t \right ] 
\\&+ \left [ c(R) + (q_{i_1} + q_{i_2} + \ldots + q_{i_r}) \right ] \\
=& \left [ \sum_{j=1}^r (k - i_j)q_{i_j} 
+ \sum_{j=0}^{r} \sum_{\substack{ t \notin R \\ t \in [ i_j + 1, i_{j+1}]} } (r-j)q_t \right ] 
\\&+  \left [ c(R) + (q_{i_1} + q_{i_2} + \ldots + q_{i_r}) \right ]\\
=& \sum_{j=1}^r \left ( q_1 + q_2 + \ldots + q_{i_j - 1} + (k - i_j + 1)q_{i_j} \right ) ~+ c(R) \\
=& ~ 2\cdot  c(R)
\end{align*}
where in the first term of these equalities, we distinguish between the cost change of all jobs not in the coalition and the cost change of the jobs in the coalition, disapearing from the game. The potential change after the first phase is simply the sum of all the changes over all machines, so $\Delta \Phi = - c(S)$.   

By the same argument, after the second phase we have $\Delta \Phi = c'(S)$. Therefore, the net change over both phases is $c'(S) - c(S)$. In conclusion, the game is a strong potential game.
\end{myproof}

%%%%%%%%%%%%%%%%%%%%%%%%%%%%%%%%%%%%%%%%%%
%*************************************************************************************************
%		Inefficiency of equilibrium
%*************************************************************************************************
%%%%%%%%%%%%%%%%%%%%%%%%%%%%%%%%%%%%%%%%%%

\section{Inefficiency of Equilibria under the \textsf{EQUI} policy}	\label{section:inefficiency}

In this section, we study the inefficiency of the game under the \textsf{EQUI} policy which is captured by the price of anarchy (PoA) and the strong price of anarchy (SPoA).  Note that the set of strong Nash equilibria is a subset of that of Nash equilibria so the SPoA is at most as large as the PoA. We state the main theorem of this section.
Whenever we bound \emph{(S)PoA} we mean that the bound applies to both the price of anarchy and the strong price of anarchy.\\

\noindent
\textbf{Summary of results on the price of anarchy:}
{ \it The game under the \textsf{EQUI} policy has the following inefficiency.
 \begin{enumerate}
   \item For identical machines, the $\textrm{(S)PoA}$ is $2 - \frac{1}{m}$. 
   \item For uniform machines, the \textrm{(S)PoA} is $ \Theta(\min\{ \log m , r \})$ where $r$ is the number of different machine's speeds in the model.
   \item For restricted identical machines, the \textrm{(S)PoA} is $\Theta(\log m)$. 
   \item For unrelated machines, the \textrm{(S)PoA} is $\Theta(m)$.  
 \end{enumerate}
}

We first give a characterization for strong Nash equilibrium in the game, which connects the equilibria to the strong ones. This characterization is useful in settling tight bounds of the strong price of anarchy in the game.

\begin{lemma}		\label{lem:strongNash}
Suppose in a Nash equilibrium there is a coalition $T$ that makes a collective move such that each job in $T$ improves strictly its cost.  Then this move preserves the number of jobs on every machine.
\end{lemma}
\begin{myproof}
For a proof by contradiction, let $\eta$ be an equilibrium that is not strong, and let $T$ be a coalition as stated in the claim. Suppose that the number of jobs on the machines is not preserved by the move of $T$.  Let $j$ be a machine that has strictly more jobs after the move, and among all jobs migrating to $j$, let $o\in T$ be the job with smallest length $p_o$. 
Let $k$ and $k'$ be the numbers of jobs on $j$ before and after the move of $T$, respectively ($k' > k$).  We claim that job $o$ could already improve its cost by unilaterally moving to $j$, contradicting that $\eta$ is a Nash equilibrium.  Consider equilibrium $\eta$, if $o$ moves to machine $j$, its cost would be:
\begin{align*}
c_o &= (k+1-w)p_{o,j} + \sum_{i : p_{i,j} < p_{o,j}} p_{i,j}\\
 &= (k-w+1)p_{o,j} + \sum_{i : p_{i,j} < p_{o,j}, i \notin T} p_{i,j} + \sum_{i : p_{i,j} < p_{o,j}, i \in T} p_{i,j}
\end{align*}
where $w$ is the number of jobs on machine $j$ in $\eta$ with length strictly less than $p_{o,j}$.

Let $w'$ be the number of jobs on machine $j$ after the move of $T$ with length strictly less than $p_{o,j}$. Since $o$ has the smallest length among all jobs migrating to $j$, $w' \leq w$. The cost of $o$ after the move of $T$ is:
$$
c'_{o} = (k' - w')p_{o,j} + \sum_{i: p_{i,j} < p_{o,j}, i \notin T} p_{i,j}
$$
We have:
\begin{align*}
c'_o - c_o &= \left [ (k' - w') - (k - w+1) \right ] p_{i,j} - \sum_{i : p_{i,j} < p_{o,j}, i \in T} p_{i,j} \\ 
&\geq (w - w') p_{i,j} - \sum_{i : p_{i,j} < p_{o,j}, i \in T} p_{i,j} \\
&\geq (w - w') p_{i,j} - (w - w') p_{i,j} =  0
\end{align*}
where the first inequality follows from $k' \geq k + 1$ and the second inequality uses $|\{i: p_{i,j} < p_{o,j}, i \in T\}|=w - w'$. Since job $o$ has incentive to cooperate and move to machine $j$, $o$ also get better off by unilaterally changing its strategy, so $\eta$ is not an equilibrium. 
\end{myproof}

%%%%%%%%%%%%%%%%%%%%%%%%%%%%%
%  Identical machines
%%%%%%%%%%%%%%%%%%%%%%%%%%%%%
\subsection{Identical machines}	\label{subsection:identical}
In case of identical machines, the analysis of the PoA is quite similar to the well-known analysis of Graham's greedy load balancing algorithm that assigns the jobs to the least load machine, processing jobs in arbitrary order, see~\cite{Graham:Bounds-for-certain-multiprocessing}.  Here we show that the (S)PoA matches exactly the approximation factor of the greedy algorithm.

\begin{proposition} \label{pro:id-poa-2-1m}
For identical machines, the $\textrm{(S)PoA}$ is $2 - \frac{1}{m}$.  Moreover, there is an instance in which all equilibria have cost at least  $(2 - \frac{2}{m}) OPT$.
\end{proposition}
\begin{myproof}
  \emph{(Upper bound)} $\quad$ First we prove that \textrm{PoA} is upper-bounded by $2-1/m$.  Let $\sigma$ be an   equilibrium and $\ell_{\max}$ be the makespan of this equilibrium.  Let $i$ be a job (with processing time $p_i$)   that has cost $\ell_{\max}$.  Hence, $p_i \leq OPT$.  Since $\sigma$ is an equilibrium, the fact that job $i$ has no   incentive to move to any other machine $j$ implies $\ell_{\max} \leq \ell_j + p_i$ for all machines $j$ different to   $\sigma(i)$, where $\ell_j$ is the load of machine $j$.  Summing up these inequalities over all machines $j$ we get   $m\ell_{\max} \leq \sum_{j=1}^m \ell_j + (m-1)p_i$.  Moreover, for any assignment of jobs to identical machines,   $\sum_{j=1}^m \ell_j \leq m OPT$. Therefore, $m\ell_{\max} \leq (2m-1)OPT$, i.e., $\textrm{PoA} \leq 2 - 1/m$.

\emph{(Lower bound)} $\quad$ Now we give an instance in which $OPT$
  equals $m$ and all equilibria have cost at least $2m - 2$. In the
  instance, there are $m$ machines and $m(m-1) + 1$ jobs in which all
  jobs have processing time 1 except one with processing time $m$. In an optimum
  assignment, the big job is scheduled on one machine and
  all $m(m-1)$ unit jobs are evenly assigned to the other machines, producing makespan $m$. We claim that in any equilibrium,
  every machine has at least $(m-1)$ unit jobs. Suppose there is
  a machine with at most $m-2$ jobs.  Since there are $m(m-1)$ jobs of unit processing time, there must be a machine $j$ with at least $m$ unit jobs in the equilibrium.  A unit job on machine $j$  has cost at least $m$ and it has incentive to move to the machine with
  less than $m-2$ jobs and get a smaller cost (at most $m-1$). This gives a contradiction and shows that any equilibrium, every machine has at least $m-1$ unit jobs. 
   Now consider the machine with the big job.  In addition this machine has at least $m-2$ unit jobs, so its load is
  at least $m + (m-2)$. Therefore, the makespan of the equilibrium is at least $(2 - 2/m)OPT$.

  Consider the schedule in which there are $(m-1)$ unit jobs
  on every machine and the job with processing time $m$ on some arbitrary machine. It is
  straightforward that this is an equilibrium. By Lemma~\ref{lem:strongNash}, this equilibrium is also a strong one. Hence,
  $\textrm{(S)PoA} \geq 2 - 1/m$.
\end{myproof}

%%%%%%%%%%%%%%%%%%%%%%%%%%%%%
%  Uniform machines
%%%%%%%%%%%%%%%%%%%%%%%%%%%%%
\subsection{Uniform machines}	\label{subsection:uniform}

For uniform machines, an upper bound $O(\log m)$ on the PoA of any deterministic policy in this machine environment
is proved by \citet{ImmorlicaLiMirrokni:Coordination-Mechanisms-for-Selfish}.
In this section, we investigate the lower bound and show that the bound $O(\log m)$ is essentially tight. 

In the following, we present a family of game instances in which the PoA, together with the SPoA, are $\Omega(\log m)$. The instances are inspired by the ones proving the lower bound of the competitive ratio of the greedy algorithm for uniform machine in~\cite{AspnesAzarFiat:On-line-routing-of-virtual}.  

\begin{figure}[htbp]
    \centering{
        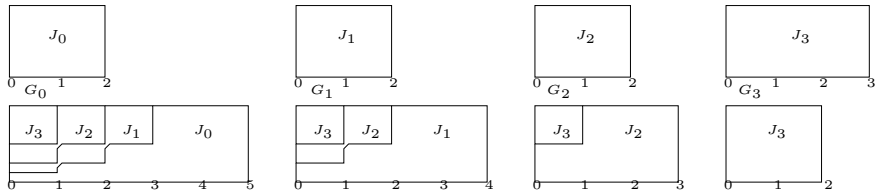
    }
    \caption{Illustration of the schedule with makespan 3 (upper part) and the strategy profile $\sigma$ (lower part) in the game instance for $k = 3$. Each machine group is represented by one of its machines.}
    \label{fig:LB-related}
  \end{figure}

\paragraph{Family of Game Instances} There are $k+1$ groups of machines $G_0, G_1, \ldots, G_k$, each machine in group $G_j$ has speed $2^{-j}$ for $0 \leq j \leq k$. Group $G_0$ has $m_0 = 1$ machine, group $G_j$ has $m_j$ machines which is recursively defined as $m_j = \sum_{t = 0}^{j-1} m_t \cdot 2^{j-t}$. Moreover, there are $k+1$ groups of jobs $J_0, J_1, \ldots, J_k$, for $0 \leq j \leq k-1$ each group $J_j$ consists of $2 m_j$ jobs of length $2^{-j}$  and group $J_k$ consists of $3 m_k$ jobs of length $2^{-k}$. The total number of machines is $m = \sum_{j=0}^{k} m_j = 1 + \frac{2}{3}(4^k - 1) + 2\cdot 4^{k-1}$, thus $k = \Omega(\log m)$. See Figure~\ref{fig:LB-related} for illustration.

Consider a schedule that is a two-to-one mapping from the job group $J_j$ to the   machine group $G_j$, for every $j<k$, and that is a three-to-one mapping from   job group $J_k$ to   machine group $G_k$.  The load on a machine in group $G_j$ for $j<k$ is 2 and each machine in $G_k$ has load 3. Hence, $OPT \leq 3$.  

Consider a schedule (strategy profile) $\sigma$ such that for every $0 \leq j \leq k$, in each machine of group $G_j$ (with speed $2^{-j}$), there are $2$ jobs of length $2^{-j}$, and for every $j<i\leq k$, there are $2^{i-j}$ jobs of length $2^{-i}$.  Each machine in group $G_j$ has $2^{k-j+1}$ jobs and has load $2 \cdot 2^{-j}/2^{-j} + \sum_{t=1}^{k-j} 2^{t} \cdot 2^{-(j+t)} / 2^{-j} = k - j + 2$, so the makespan of this schedule is $k+2$. 
We claim that this strategy profile is a Nash equilibrium, moreover it is a strong one.

\begin{lemma}  
The strategy profile $\sigma$ is a Nash equilibrium.
\end{lemma}
\begin{myproof}
First, we show that, in strategy profile $\sigma$, the cost of a job in $J_j$ is equal to $k-j+2$. Fix a machine in $G_t$. We are only interested in case $t \leq j$ since in $\sigma$, no job in $J_j$ is assigned to a machine of group $G_t$ with $t > j$. On this machine, there are exactly $2^{j-t+1}$ jobs with processing time at least $2^{-j}$. So, the cost of a job in $J_j$ scheduled on this machine is:
\begin{align*}
&\frac{1}{2^{-t}} \left [ \left ( 2^{k-t}\cdot 2^{-k} + 2^{(k-1)-t}\cdot 2^{-(k-1)} + \ldots + 2^{(j+1)-t}\cdot 2^{-(j+1)} \right ) + 2^{j-t+1}\cdot 2^{-j} \right ] \\
& =\: k + 2 - j.
\end{align*}
  
Now we argue that $\sigma$ is an equilibrium.  Suppose that a job $i$ in $J_j$ moves from its current machine to a machine of group $G_t$. If $j \leq t$, $i$ has the greatest length among all jobs assigned to this new machine, so the new cost of $i$ is the new load of the machine which is $(k-t+2) + 2^{-j}/2^{-t} > k-j+2$. If $j > t$ then there are $(2^{j-t+1} +1)$ jobs with length at least $2^{-j}$ on $i$'s new machine. Hence, the new cost of $i$ is:
\begin{align*}
&\frac{1}{2^{-t}} \left [ \left ( 2^{k-t}\cdot 2^{-k} + 2^{(k-1)-t}\cdot 2^{-(k-1)} + \ldots + 2^{(j+1)-t}\cdot 2^{-(j+1)} \right ) + (2^{j-t+1}+1)\cdot 2^{-j} \right ]\\ 
 &>\: k + 2 - j.
\end{align*}
Therefore, no job can improve its cost by changing its strategy.
\end{myproof}

Using Lemma \ref{lem:strongNash}, we show that $\sigma$ is indeed a strong equilibrium.

\begin{lemma}	\label{lemma:strongNash}
Strategy profile $\sigma$ is a strong Nash equilibrium.
\end{lemma}
\begin{myproof}
Suppose $\sigma$ is not a strong Nash equilibrium, then there exists a coalition $T$ such that all jobs in $T$ strictly decrease their costs and after the move of $T$, all machines have the same number of jobs as in $\sigma$ (by Lemma~\ref{lem:strongNash}). Observe that the cost of a job in $J_k$ (with the least length among all jobs in the instance) depends only on the number of jobs scheduled on its machine. With such a move of $T$, if there are some jobs in $J_k$ involved in the coalition, none of them can strictly decrease its cost. Hence, $T \cap J_k = \emptyset$. Consider jobs in $J_{k-1}$. Since jobs in $J_k$ stay in their machines and they incur the same load 1 on each machine, the cost of a job in $J_{k-1}$, if it involves in $T$, depends only on the number of jobs which are not in $J_k$ and are scheduled on its new machine. However, this number is preserved after the move of $T$ (by Lemma~\ref{lem:strongNash} and $T \cap J_k = \emptyset$), so the cost of a job in $J_{k-1}$ stays the same, i.e., the job has no incentive to involve in $T$. The argument holds for groups of jobs $J_{k-2}, \ldots, J_0$. Therefore, $T = \emptyset$ meaning that $\sigma$ is a strong equilibrium. 
\end{myproof}

The previous lemmas imply that for uniform machines, the \textrm{PoA} of \textsf{EQUI} is $\Omega(\log m)$.
\begin{theorem}
For uniform machines, the \textrm{(S)PoA} of \textsf{EQUI} is $\Theta(\log m)$.
\end{theorem}

%%%%%%%%%%%%%%%%%%%%%%%%%%%%%
%  Restricted identical machines
%%%%%%%%%%%%%%%%%%%%%%%%%%%%%
\subsection{Restricted Identical Machines}	\label{subsection:restricted}
 
The upper bound of the price of anarchy of \textsf{EQUI} on restricted identical machines follows immediately by \citet{ImmorlicaLiMirrokni:Coordination-Mechanisms-for-Selfish}. In \cite{ImmorlicaLiMirrokni:Coordination-Mechanisms-for-Selfish}, an instance  was given which shows that any deterministic non-preemptive coordination mechanism has PoA $\Omega(\log m)$. However, \textsf{EQUI} is a preemptive policy, and the instance cannot be adapted. In this section, we show that the price of anarchy of the  \textsf{EQUI} policy on restricted identical machines is also $\Omega(\log m)$ using another instance.

\begin{theorem} \label{thm:res-poa-logm}
For restricted identical machines, the (S)PoA is $\Theta(\log m)$.
\end{theorem}
\begin{myproof}
The upper bound follows from \cite{ImmorlicaLiMirrokni:Coordination-Mechanisms-for-Selfish}. We show now the lower bound.
We adapt a game instance from the proof of Lemma~\ref{lemma:strongNash}.  Let $(m_j)_{j=0}^k$ be a sequence defined as $m_0 = 1, m_1 = 2$ and $m_j = m_0 + \ldots + m_{j-1}$, i.e., $m_j = 3 \cdot 2^{j-2}$ for every $j \geq 2$. Let $m = \sum_{j=0}^k m_j = 3 \cdot 2^{k-1}$. Hence $k = \Omega(\log m)$. 

In the instance, there are $m$ machines which are divided into $k+1$ groups $G_0,\ldots,G_k$ where group $G_j$ consists of $m_j$ machines.  There are also $k+1$ job groups $J_0, J_1, \ldots, J_k$ where group $J_j$ contains $3\cdot2^j m_j$ jobs of processing time $2^{-j}$. 

We first describe a schedule $\mu$ which will be proved to be a Nash equilibrium. On each machine in group $G_j $ for $0 \leq j \leq k$, there are $2^{j+1}$ jobs of length $2^{-j}$ and for every $j<i\leq k$, there are $2^i$ jobs of length $2^{-i}$. The strategy set of each job is the following. Jobs in group $J_j$ can be scheduled on all $m_j$ machines of group $G_j$. Moreover, a job in $J_j$ can be additionally scheduled on its current machine in $\mu$. 

We claim that $\mu$ is a equilibrium. Observe that on each machine of group $G_j$, there are exactly $2^{i+1}$ jobs of processing time at least $2^{-i}$ for all $j \leq i \leq k$ and the total load of jobs with processing time strictly smaller than $2^{-i}$ (on the machine) is $k - i$.  Thus, the cost of each job in group $J_i$ is $k-i+2$ in $\mu$ and if a job switches the strategy, its cost would be strictly greater than $k-j+2$.  In addition, using Lemma~\ref{lem:strongNash} and by the same argument as in Lemma~\ref{lemma:strongNash}, we have that this equilibrium is indeed a strong one.

If we schedule evenly all jobs of group $J_j$ on $m_j$ machines of $G_j$ for $0 \leq j \leq k$ then the makespan is bounded by 3, so $OPT \leq 3$. The makespan of the strong equilibrium above is $k+1$, which gives the (S)PoA is at least $(k+1)/3 = \Omega(\log m)$.
\end{myproof}

%%%%%%%%%%%%%%%%%%%%%%%%%%%%%
%  Unrelated machines
%%%%%%%%%%%%%%%%%%%%%%%%%%%%%

\subsection{Unrelated Machines}	\label{subsection:unrelated} 
In this section, we prove that the PoA of the game under the \textsf{EQUI} policy is upper bounded by $2m$.  Interestingly, without any knowledge of jobs' characteristics, the inefficiency of \textsf{EQUI} -- a non-clairvoyant policy -- is  the same up to a constant compared to that of \textsf{SPT} -- the best strongly local policy with price of anarchy $\Theta(m)$.

\begin{theorem}
For unrelated machines, the price of anarchy of policy \textsf{EQUI} is at most $2m$.
\end{theorem}
\begin{myproof}
For job $i$, let $q_i$ be the smallest processing time of $i$ among all machines, i.e., $q_i := \min_j p_{i,j}$ and let $Q(i)$ be the machine $j$ minimizing $p_{i,j}$.  Without loss of generality we assume that jobs are indexed  such that $q_1 \leq q_2 \leq \ldots \leq q_n$.  Note that $\sum_{i=1}^n q_i \leq m \cdot OPT$, where $OPT$ is the optimal makespan, as usual.  First, we claim the following lemma.

We claim that  \label{lem:unrel-cost}
In any Nash equilibrium, the cost $c_i$ of job $i$ is at most 
\begin{equation}                                 \label{eq:qi}
  2q_1 + \ldots + 2q_{i-1} + (n-i+1)q_i.
\end{equation}
The theorem would follow from the claim by the following argument.
Since the expression (\ref{eq:qi}) is increasing in $i$ and at $i = n$ this term is $2\sum_{i=1}^n q_i \leq 2m\cdot OPT$, the  cost of each job in an equilibrium is bounded by $2m \cdot OPT$, so the price of anarchy is at most $2m$.  

The proof of the claim is by induction on $i$.  The cost of job $1$ on machine $Q(1)$ would be at most $n q_1$, simply because there are at most $n$ jobs on this machine.  Therefore the cost of job $1$ in the Nash equilibrium is also at most $n q_1$.  Assume the induction hypothesis holds until index $i-1$.  Consider job $i$.  Since the strategy profile is a Nash equilibrium, $i$'s current cost is at most its cost if moving to machine $Q(i)$.  We distinguish different cases. In these cases, denote $c'_i$ as the new cost of $i$ if it moves to machine $Q(i)$
\begin{enumerate}
\item Case all jobs $t$ scheduled on machine $Q(i)$ satisfy $t>i$.

        This case is very similar to the basis case.  There are at most $n-i$ jobs on machine $Q(i)$, beside $i$.  The completion time of job $i$ is then at most $(n-i+1)q_i$ which is upper bounded by (\ref{eq:qi}).  For the remaining cases, we assume that there is a job $i'<i$ scheduled on $Q(i)$.

\item Case there is a job $t < i$ on machine $Q(i)$ such that $p_{t,Q(i)} \geq p_{i,Q(i)} (= q_i)$.
	
	Since $p_{t,Q(i)} \geq q_i$, the new cost of job $i$ is not more than the new cost of job $t$.  Moreover, the new cost of job $t$ is increased by exactly $q_i$, so the new cost of $i$ is bounded by
	\begin{eqnarray*}
	c'_i &\leq& c_t + q_i \\
	& \leq  & 2 q_1 + \ldots + 2 q_{t-1}  + (n-t+1)q_{t} + q_i \\ 
	& =   & 2 q_1 + \ldots + 2 q_{t-1} + 2(i-t)q_t + (n-2i+ t + 1)q_{t} + q_i \\ 
	&\leq & 2 q_1 + \ldots + 2 q_{t-1}+ 2q_{t} + \ldots + 2 q_{i-1} + (n-i+1)q_{i},
	\end{eqnarray*}
where the first inequality uses the induction hypothesis and the last inequality is due to $t < i$ and $q_t \leq q_{t+1} \leq \ldots \leq q_i$.

\item Case every job $t$ scheduled on machine $Q(i)$ with $p_{t,Q(i)} \geq q_i$ satisfies $t\geq i$.

        Since we are not in the first two cases, there is a job $t<i$ on machine $Q(i)$ with $p_{t,Q(i)} < q_i$.  Let $i'$ be the job of greatest index among all jobs scheduled on $Q(i)$ with smaller processing time than $q_i$.  All jobs $t$ scheduled on $Q(i)$ and having smaller processing time than that of $i$, also have smaller index because $q_t \leq p_{t,Q(i)} \leq q_i$.  Therefore $i'$ is precisely the last job to complete before $i$.  At the completion time of $i'$ there are still $q_i-p_{i',Q(i)}\leq q_i-q_{i'}$ units of $i$ to be processed.  By the case assumption, there are at most $(n-i)$ jobs with processing time greater than that of $i$.  Therefore the new cost of $i$ is at most
	\begin{eqnarray*}
	c'_i &= &c_{i'} + (n-i+1)(q_i-q_{i'}) \\
	 &\leq    &2 q_1 + \ldots + 2 q_{i'-1}+(n-i'+1)q_{i'} + (n-i+1)(q_i-q_{i'}) \\ 
	&=   &2 q_1 + \ldots + 2 q_{i'-1}+(i-i')q_{i'} + (n-i+1)q_i  \\ 
	&\leq&2 q_1 + \ldots + 2 q_{i'-1}+ ( q_{i'} + \ldots + q_{i-1} )+ (n-i+1)q_i  \\ 
	&\leq&2 q_1 + \ldots + 2 q_{i-1} +(n-i+1)q_i 
	\end{eqnarray*}
where the first inequality uses the induction hypothesis and the third inequality is due to the monotonicity of the sequence $(q_j)_{j=1}^n$.
\end{enumerate}
This completes the proof of the claim, and therefore of the theorem.
\end{myproof}

We provide a game instance showing that the upper bound analyzed above is tight. The instance is inspired by the work of \citet{AzarJainMirrokni:Almost-Optimal-Coordination}. In the following lemma, we prove the lower bound of the PoA of the game under the \textsf{EQUI} policy. 

\begin{lemma}\label{lem:unrel-poa-m14}
The (strong) price of anarchy of \textsf{EQUI} is at least $(m+1)/4$.
\end{lemma}
\begin{myproof}
Let $n_j := \frac{2 (m-1)!}{(j-1)!}$ and $n := \sum_{j=1}^m n_j$. Consider the set of $m$ machines and $m$ groups of jobs $J_1, J_2, \ldots, J_m$. In group $J_j$ $(1 \leq j \leq m-1)$, there are $n_j$ jobs that can be scheduled on machine $j$ or $j+1$ except the last group ($J_m$) which has a single job that can be only scheduled on machine $m$. Each job in group $J_j$ $(1 \leq j \leq m-1)$ has processing time $p_{j,j} = \frac{(j-1)!}{(m-1)!} = \frac{2}{n_j}$ on machine $j$ and has processing time $p_{j,j+1} = \frac{j!}{2(m-1)!} = \frac{1}{n_{j+1}}$ on machine $j+1$. The job in $J_m$ has processing time $p_{m,m} = 1$ on machine $m$. 

Consider the strategy profile in which half of the jobs in $J_j$ $(1 \leq j \leq m-1)$ are scheduled on machine $j$ and the other half are scheduled on machine $j+1$ (jobs in $J_m$ are scheduled on machine $m$). We claim that this strategy profile is a Nash equilibrium. Note that the cost of jobs in the same group and scheduled on the same machine are the same. The cost of each job in group $J_j$ on machine $j$ is the load of  the machine, because its processing time is greater than that of jobs in group $J_{j-1}$ on machine $m$, and this load equals $\frac{n_{j-1}}{2}p_{j-1,j} + \frac{n_j}{2}p_{j,j} = \frac{j-1}{2} + 1 = \frac{j+1}{2}$. Each job in group $J_j$ has smaller processing time than that of each job in group $J_{j+1}$  on machine $j+1$, thus the cost of the former is $\frac{n_j + n_{j+1}}{2} p_{j,j+1} = \frac{j+1}{2}$. Hence, no job in group $J_j$ ($1 \leq j \leq m-1$) has an incentive to move and the job in group $J_m$ cannot switch its strategy. Therefore, the strategy profile is an equilibrium. 

Moreover, we prove that this equilibrium is indeed a strong one. Suppose that it is not a strong equilibrium, i.e., there is a coalition $S$ such that all jobs in $S$ can strictly decrease their cost. Again, by Lemma~\ref{lem:strongNash}, the number of jobs on each machine remains the same after the move of $S$. We call a job in group $J_j$ \emph{moving up} if it moves from machine $j$ to $j+1$ and \emph{moving down} if it moves from machine $j+1$ to $j$. First, we claim that no job has an incentive to move up. If a job in group $J_j$ moves up, as only jobs in $J_j$ and $J_{j+1}$ can use machine $j+1$ and $p_{j,j+1} < p_{j+1, j+1}$, its new cost would be $p_{j,j+1} \cdot (n_j + n_{j+1})$ which equals its old cost. Hence, no one can strictly decrease its cost by moving up. Among all jobs in $S$, consider the one who moves down to the machine $j^*$ of smallest index. By the choice of $j^*$, there is no job moving down from machine $j^*$ and as claimed above, no job moving up from $j^*$. Hence, the job moving to machine $j^*$ cannot strictly decrease its cost -- that contradicts to the assumption that all jobs in $S$ strictly get better off. Therefore, the equilibrium is a strong one.

Consider a schedule in which jobs in group $J_j$ ($1 \leq j \leq m$) are assigned to machine $j$ and this schedule has makespan 2, hence $OPT \leq 2$. The makespan of the above (strong) Nash equilibrium is the load on machine $m$, that is equal to $(m+1)/2$. Then, the (strong) price of anarchy is at least $(m+1)/4$.
\end{myproof}

%
%%%%%%%%%%%%%%%%%%%%%%%%%%%%%%%%%%%%%%%%%
%%******************************************************************************************
%%		Conclusion + Open questions
%%******************************************************************************************
%%%%%%%%%%%%%%%%%%%%%%%%%%%%%%%%%%%%%%%%%

\section{Conclusion and Open questions}

In this paper, we studied coordination mechanisms under non-clairvoyant policies. We first studied whether some policies are admissible -- which is the first property that we expect from a policy. We studied in detail the existence of Nash equilibrium under the \textsf{RANDOM} and the \textsf{EQUI} policies. Next, we analyzed the inefficiency (PoA) of the \textsf{EQUI} policy and showed that the knowledge of the agents processing times is not really necessary, since \textsf{EQUI} behaves nearly as good as the best known strongly local policy \textsf{SPT}. One more advantage is that there is no need to implement \textsf{EQUI} policy (if using it) since this popular policy exists in many operating systems.

An interesting open question is to answer (prove or disprove) whether the gap of the PoA between strongly local and local policies can be closed. Does there exist a (preemptive, randomized) strongly local policies with the PoA poly-logarithmic on $m$?  \citet{AzarJainMirrokni:Almost-Optimal-Coordination} proved that with an additional condition, this gap is closed. Can we bypass this condition? Besides, does there exist a truthful coordination mechanism based on strongly local policy with PoA as $o(m)$~?

Another interesting open problem is the speed of convergence to approximate a Nash equilibrium for \textsf{RANDOM} and \textsf{EQUI} in the machines environment where equilibrium is guaranteed to exist. 

\paragraph{Acknowledgments} \: We would like to thank Adi Ros\'en for helpful discussions.

\bibliographystyle{plainnat}
\bibliography{gametheory}

\end{document}

%% file: relatedLB.pdf_t
\begin{picture}(0,0)%
\includegraphics{relatedLB.pdf}%
\end{picture}%
\setlength{\unitlength}{829sp}%
\begingroup\makeatletter\ifx\SetFigFont\undefined%
\gdef\SetFigFont#1#2#3#4#5{%
  \reset@font\fontsize{#1}{#2pt}%
  \fontfamily{#3}\fontseries{#4}\fontshape{#5}%
  \selectfont}%
\fi\endgroup%
\begin{picture}(25854,5527)(3486,-4248)
\put(14931,-1161){\makebox(0,0)[lb]{\smash{{\SetFigFont{5}{6.0}{\rmdefault}{\mddefault}{\updefault}{\color[rgb]{0,0,0}2}%
}}}}
\put(4001,-2590){\makebox(0,0)[lb]{\smash{{\SetFigFont{5}{6.0}{\rmdefault}{\mddefault}{\updefault}{\color[rgb]{0,0,0}$J_3$}%
}}}}
\put(6930,-2590){\makebox(0,0)[lb]{\smash{{\SetFigFont{5}{6.0}{\rmdefault}{\mddefault}{\updefault}{\color[rgb]{0,0,0}$J_1$}%
}}}}
\put(5501,-2590){\makebox(0,0)[lb]{\smash{{\SetFigFont{5}{6.0}{\rmdefault}{\mddefault}{\updefault}{\color[rgb]{0,0,0}$J_2$}%
}}}}
\put(4001,-1375){\makebox(0,0)[lb]{\smash{{\SetFigFont{5}{6.0}{\rmdefault}{\mddefault}{\updefault}{\color[rgb]{0,0,0}$G_0$}%
}}}}
\put(4930,-4233){\makebox(0,0)[lb]{\smash{{\SetFigFont{5}{6.0}{\rmdefault}{\mddefault}{\updefault}{\color[rgb]{0,0,0}1}%
}}}}
\put(6358,-4233){\makebox(0,0)[lb]{\smash{{\SetFigFont{5}{6.0}{\rmdefault}{\mddefault}{\updefault}{\color[rgb]{0,0,0}2}%
}}}}
\put(7787,-4233){\makebox(0,0)[lb]{\smash{{\SetFigFont{5}{6.0}{\rmdefault}{\mddefault}{\updefault}{\color[rgb]{0,0,0}3}%
}}}}
\put(9216,-4233){\makebox(0,0)[lb]{\smash{{\SetFigFont{5}{6.0}{\rmdefault}{\mddefault}{\updefault}{\color[rgb]{0,0,0}4}%
}}}}
\put(12645,-2590){\makebox(0,0)[lb]{\smash{{\SetFigFont{5}{6.0}{\rmdefault}{\mddefault}{\updefault}{\color[rgb]{0,0,0}$J_3$}%
}}}}
\put(14074,-2590){\makebox(0,0)[lb]{\smash{{\SetFigFont{5}{6.0}{\rmdefault}{\mddefault}{\updefault}{\color[rgb]{0,0,0}$J_2$}%
}}}}
\put(12073,-4233){\makebox(0,0)[lb]{\smash{{\SetFigFont{5}{6.0}{\rmdefault}{\mddefault}{\updefault}{\color[rgb]{0,0,0}0}%
}}}}
\put(13502,-4233){\makebox(0,0)[lb]{\smash{{\SetFigFont{5}{6.0}{\rmdefault}{\mddefault}{\updefault}{\color[rgb]{0,0,0}1}%
}}}}
\put(15002,-4233){\makebox(0,0)[lb]{\smash{{\SetFigFont{5}{6.0}{\rmdefault}{\mddefault}{\updefault}{\color[rgb]{0,0,0}2}%
}}}}
\put(16431,-4233){\makebox(0,0)[lb]{\smash{{\SetFigFont{5}{6.0}{\rmdefault}{\mddefault}{\updefault}{\color[rgb]{0,0,0}3}%
}}}}
\put(19789,-2590){\makebox(0,0)[lb]{\smash{{\SetFigFont{5}{6.0}{\rmdefault}{\mddefault}{\updefault}{\color[rgb]{0,0,0}$J_3$}%
}}}}
\put(19217,-4233){\makebox(0,0)[lb]{\smash{{\SetFigFont{5}{6.0}{\rmdefault}{\mddefault}{\updefault}{\color[rgb]{0,0,0}0}%
}}}}
\put(20646,-4233){\makebox(0,0)[lb]{\smash{{\SetFigFont{5}{6.0}{\rmdefault}{\mddefault}{\updefault}{\color[rgb]{0,0,0}1}%
}}}}
\put(22146,-4233){\makebox(0,0)[lb]{\smash{{\SetFigFont{5}{6.0}{\rmdefault}{\mddefault}{\updefault}{\color[rgb]{0,0,0}2}%
}}}}
\put(13359,268){\makebox(0,0)[lb]{\smash{{\SetFigFont{5}{6.0}{\rmdefault}{\mddefault}{\updefault}{\color[rgb]{0,0,0}$J_1$}%
}}}}
\put(9073,-2590){\makebox(0,0)[lb]{\smash{{\SetFigFont{5}{6.0}{\rmdefault}{\mddefault}{\updefault}{\color[rgb]{0,0,0}$J_0$}%
}}}}
\put(25361,-1375){\makebox(0,0)[lb]{\smash{{\SetFigFont{5}{6.0}{\rmdefault}{\mddefault}{\updefault}{\color[rgb]{0,0,0}$G_3$}%
}}}}
\put(26432,-4233){\makebox(0,0)[lb]{\smash{{\SetFigFont{5}{6.0}{\rmdefault}{\mddefault}{\updefault}{\color[rgb]{0,0,0}1}%
}}}}
\put(27933,-4233){\makebox(0,0)[lb]{\smash{{\SetFigFont{5}{6.0}{\rmdefault}{\mddefault}{\updefault}{\color[rgb]{0,0,0}2}%
}}}}
\put(19646,-1375){\makebox(0,0)[lb]{\smash{{\SetFigFont{5}{6.0}{\rmdefault}{\mddefault}{\updefault}{\color[rgb]{0,0,0}$G_2$}%
}}}}
\put(20503,268){\makebox(0,0)[lb]{\smash{{\SetFigFont{5}{6.0}{\rmdefault}{\mddefault}{\updefault}{\color[rgb]{0,0,0}$J_2$}%
}}}}
\put(23503,-4233){\makebox(0,0)[lb]{\smash{{\SetFigFont{5}{6.0}{\rmdefault}{\mddefault}{\updefault}{\color[rgb]{0,0,0}3}%
}}}}
\put(17788,-4233){\makebox(0,0)[lb]{\smash{{\SetFigFont{5}{6.0}{\rmdefault}{\mddefault}{\updefault}{\color[rgb]{0,0,0}4}%
}}}}
\put(4715,268){\makebox(0,0)[lb]{\smash{{\SetFigFont{5}{6.0}{\rmdefault}{\mddefault}{\updefault}{\color[rgb]{0,0,0}$J_0$}%
}}}}
\put(10645,-4233){\makebox(0,0)[lb]{\smash{{\SetFigFont{5}{6.0}{\rmdefault}{\mddefault}{\updefault}{\color[rgb]{0,0,0}5}%
}}}}
\put(16217,-2590){\makebox(0,0)[lb]{\smash{{\SetFigFont{5}{6.0}{\rmdefault}{\mddefault}{\updefault}{\color[rgb]{0,0,0}$J_1$}%
}}}}
\put(21932,-2590){\makebox(0,0)[lb]{\smash{{\SetFigFont{5}{6.0}{\rmdefault}{\mddefault}{\updefault}{\color[rgb]{0,0,0}$J_2$}%
}}}}
\put(26218,-2590){\makebox(0,0)[lb]{\smash{{\SetFigFont{5}{6.0}{\rmdefault}{\mddefault}{\updefault}{\color[rgb]{0,0,0}$J_3$}%
}}}}
\put(26932,268){\makebox(0,0)[lb]{\smash{{\SetFigFont{5}{6.0}{\rmdefault}{\mddefault}{\updefault}{\color[rgb]{0,0,0}$J_3$}%
}}}}
\put(24932,-4233){\makebox(0,0)[lb]{\smash{{\SetFigFont{5}{6.0}{\rmdefault}{\mddefault}{\updefault}{\color[rgb]{0,0,0}0}%
}}}}
\put(3501,-4233){\makebox(0,0)[lb]{\smash{{\SetFigFont{5}{6.0}{\rmdefault}{\mddefault}{\updefault}{\color[rgb]{0,0,0}0}%
}}}}
\put(3501,-1161){\makebox(0,0)[lb]{\smash{{\SetFigFont{5}{6.0}{\rmdefault}{\mddefault}{\updefault}{\color[rgb]{0,0,0}0}%
}}}}
\put(5001,-1161){\makebox(0,0)[lb]{\smash{{\SetFigFont{5}{6.0}{\rmdefault}{\mddefault}{\updefault}{\color[rgb]{0,0,0}1}%
}}}}
\put(6358,-1161){\makebox(0,0)[lb]{\smash{{\SetFigFont{5}{6.0}{\rmdefault}{\mddefault}{\updefault}{\color[rgb]{0,0,0}2}%
}}}}
\put(12073,-1161){\makebox(0,0)[lb]{\smash{{\SetFigFont{5}{6.0}{\rmdefault}{\mddefault}{\updefault}{\color[rgb]{0,0,0}0}%
}}}}
\put(13502,-1161){\makebox(0,0)[lb]{\smash{{\SetFigFont{5}{6.0}{\rmdefault}{\mddefault}{\updefault}{\color[rgb]{0,0,0}1}%
}}}}
\put(12574,-1375){\makebox(0,0)[lb]{\smash{{\SetFigFont{5}{6.0}{\rmdefault}{\mddefault}{\updefault}{\color[rgb]{0,0,0}$G_1$}%
}}}}
\put(19217,-1161){\makebox(0,0)[lb]{\smash{{\SetFigFont{5}{6.0}{\rmdefault}{\mddefault}{\updefault}{\color[rgb]{0,0,0}0}%
}}}}
\put(20646,-1161){\makebox(0,0)[lb]{\smash{{\SetFigFont{5}{6.0}{\rmdefault}{\mddefault}{\updefault}{\color[rgb]{0,0,0}1}%
}}}}
\put(22075,-1161){\makebox(0,0)[lb]{\smash{{\SetFigFont{5}{6.0}{\rmdefault}{\mddefault}{\updefault}{\color[rgb]{0,0,0}2}%
}}}}
\put(24932,-1161){\makebox(0,0)[lb]{\smash{{\SetFigFont{5}{6.0}{\rmdefault}{\mddefault}{\updefault}{\color[rgb]{0,0,0}0}%
}}}}
\put(26361,-1161){\makebox(0,0)[lb]{\smash{{\SetFigFont{5}{6.0}{\rmdefault}{\mddefault}{\updefault}{\color[rgb]{0,0,0}1}%
}}}}
\put(27790,-1161){\makebox(0,0)[lb]{\smash{{\SetFigFont{5}{6.0}{\rmdefault}{\mddefault}{\updefault}{\color[rgb]{0,0,0}2}%
}}}}
\put(29218,-1161){\makebox(0,0)[lb]{\smash{{\SetFigFont{5}{6.0}{\rmdefault}{\mddefault}{\updefault}{\color[rgb]{0,0,0}3}%
}}}}
\end{picture}%